# Controlling Electron-Beam Emittance Partitioning for Future X-Ray Light Sources

Nikolai Yampolsky[1], Bruce Carlsten[1], Robert Ryne[2], Kip Bishofberger[1], Steven Russell[1], and Alex Dragt[3]

[1]*Los Alamos National Laboratory, Los Alamos, New Mexico 87545, USA*
[2]*Lawrence Berkeley National Laboratory, Berkeley, California 94720, USA*
[3]*University of Maryland, College Park, Maryland 20742, USA*

Motivated by the emittance requirements for future light sources, we show how longitudinal-transverse correlations can be introduced to create beams with large emittance asymmetry. This concept generalizes a key aspect of a Flat Beam Transform, which introduces initial correlations among the transverse planes, to systems with initial longitudinal-transverse correlations. We present an eigen-emittance formalism to analyze such systems. We illustrate the approach by analyzing an electron beam emitted from a photoinjector utilizing a laser with a tilted pulse front. This approach can be used to design beam delivery systems that achieve extraordinarily transversely bright electron beams.

Free-electron lasers (FELs) are presently the only sources of high-brightness hard X-ray pulses. The recent success of the Linac Coherent Light Source (LCLS) [1] opens new frontiers for fundamental studies in physics, chemistry, biology, and material science. Designing FELs producing higher photon energies (*i.e.*, higher than the 8-keV produced at LCLS) is a challenging task because of stringent constraints on the normalized rms transverse beam emittance, $\varepsilon_{nx,ny} \leq \beta\gamma\lambda_{X-ray}/4\pi$, where $\varepsilon_{nx}^2 = \gamma^2\beta^2(<x^2><x'^2>-<xx'>^2)$, and $x' = dx/dz$. In principle, this limitation can be overcome by increasing the beam energy. However, this approach is not efficient for producing photons having energies higher than 10 keV due to an increase of the beam energy diffusion in the wiggler from quantum fluctuations of single-particle synchrotron radiation. This effect results in the fundamental limit on the lowest achievable radiation wavelength in FELs, $\lambda_{X-ray} \propto \varepsilon_n^{16/15}$ [2]. Therefore, hard X-ray pulses can be generated only upon availability of transversely bright electron beams. Moreover, the availability of low-emittance beams will allow FELs to operate at lower beam energy, significantly reducing the linac cost, which is the dominant cost of an FEL facility.

Beam energy diffusion due to quantum fluctuations of incoherent synchrotron radiation in a wiggler is expected to play a significant role in a recently proposed 50 keV ($\lambda_{X-ray} = 0.25 \overset{\circ}{A}$) hard X-ray FEL facility for Matter and Radiation in Extremes (MaRIE) [3]. In order to overcome this limitation, the FEL is expected to operate with bunches having 3.4 kA peak current and 20 GeV energy. Detailed 3D numerical simulations predict that the normalized transverse emittance needs to be smaller than 0.14 μm in order to provide significant matching between the radiation and the beam transverse phase spaces. Standard transverse emittance scaling from state-of-the-art high-brightness photoinjectors is $\varepsilon_n \sim 1\,\mu m\,(q/nC)^{1/2}$, where $q$ is the bunch charge [4]. Using this scaling, a transverse emittance of 0.14 μm would indicate a maximum bunch charge of only 20 pC. To reach a peak current of 3.4 kA, the bunch would need to be highly compressed to 12 fs (about 3.4 μm). This may not be possible with a simple extrapolation of current technology, because it requires operating compression chicanes in an untested regime of coherent synchrotron radiation. Alternatively, a 500-pC bunch has to be compressed to only 150 fs to reach the peak current of 3.4 kA, consistent with demonstrated performance.

The constraint on the beam's normalized longitudinal emittance is much more relaxed compared to the limit on the transverse emittance. Assuming a 150-fsec bunch with typical relative energy spread $\Delta\gamma/\gamma \sim \rho \sim 10^{-4}$, the normalized longitudinal emittance, $\varepsilon_{nz}^2 = \gamma^2\beta^2(<c^2\delta t^2><(\delta\gamma/(\gamma\beta))^2> - <c\delta t(\delta\gamma/(\gamma\beta))>^2)$ can be as high as 180 μm.

The required emittances, $\varepsilon_{nx}/\varepsilon_{ny}/\varepsilon_{nz} = 0.14/0.14/180$ μm, for this type of X-ray FEL, lead to a total 6D phase space volume of about 3.5 μm³. This requirement on the volume is easily met by existing high-brightness photoinjectors, which have typical emittances $\varepsilon_{nx}/\varepsilon_{ny}/\varepsilon_{nz} = 0.7/0.7/1.4$ μm for a 500-pC bunch, corresponding to a volume of 0.7 μm³. However, the natural *emittance partitioning* into horizontal and longitudinal components does not meet the requirement for this type of X-ray FEL. A solution is to modify the *eigen-emittances* of a beam when it is created by introducing correlations among the phase planes of the initial distribution, then to manipulate it to remove correlations and exchange the eigen-emittances among the phase planes. This has the effect of partitioning the eigen-emittances (more precisely, targeting each to a specific phase plane) to meet emittance requirements.

There has been compelling recent work on new optics schemes that are based on manipulating emittances among the beam phase space planes. Specifically, it has been shown possible to change the ratio of two transverse emittances while keeping the transverse phase space volume intact (through a scheme called a Flat-Beam Transform (FBT) [5-8]) and to swap the longitudinal emittance with one of the transverse emittances (through a scheme called an Emittance EXchanger (EEX) [9,10]). Using these two optics tools, Kim proposed [11] to start with a temporally short pulse having very low longitudinal emittance and to apply first a FBT and then an EEX to achieve the required low transverse emittances. Consider a hypothetical case of a photoinjector driven by a 200 fs laser with initial emittances of 2.1/2.1/0.14 μm. The FBT involves applying magnetic field to the cathode to change two eigen-emittances (increasing one and decreasing the other) that will eventually become the two transverse 1D emittances downstream. This conversion of eigen-emittances into 1D emittances is achieved by two skew quadrupoles to remove correlations between the transverse planes. Depending on the design, it is possible in this example to achieve emittances of 0.14/35/0.14 µm after the FBT. The EEX would then swap the vertical and longitudinal emittances, leading to a 0.14/0.14/35 µm emittances. Unfortunately, photoinjector design does not scale linearly enough to produce the initial emittances of the hypothetical case mentioned above.

The purpose of this Letter is to describe a new technique that will allow photoinjector designers and designers of beam delivery systems to analyze and control the emittance partitioning to optimize FEL performance. This technique is based on the fact that each electron bunch has three ensemble-averaged quantities, known as the "eigen-emittances" [12], which are invariant (up to a reordering) under linear Hamiltonian dynamics. The eigen-emittances are derived from the beam 6x6 second moment matrix, $\Sigma$, when *canonical* variables are used. When $\Sigma$ is 2x2 block diagonal, the eigen-emittances are identical to the canonical 1D rms emittances, $\varepsilon_x^2 = <x^2><p_x^2> - <xp_x>^2$, *etc*. Though the eigen-emittances are invariant under linear transport, their initial values can be changed in a number of ways. Frequently this results in correlations in the beam matrix $\Sigma$ at the cathode surface. For example, since the canonical momenta are related to the mechanical momenta by, $p=p_{mech}+eA$ (where $e$ is the electron charge and $A$ is the vector potential), immersing a cathode in a solenoidal magnetic field (as in an FBT) produces $x$-$p_y$ and $y$-$p_x$ correlations in $\Sigma$ at the cathode surface and changes two of the eigen-emittances. Later we will exhibit another approach to change the eigen-emittances at the cathode surface that involves introducing transverse-temporal correlations in $\Sigma$. The key point, in general, is that, by knowing the desired 1D emittances, one can attempt to create a beam with these eigen-emittances, then design a beamline to both remove correlations and exchange the eigen-emittances appropriately among the phase planes. In this way a designer can control the values of the final 1D emittances and their partitioning among the phase planes.

To start our analysis we describe the electron bunch as an ensemble of electrons in 6D phase space. For each electron, let $\zeta=(q_1,p_1,q_2,p_2,q_3,p_3)$ denote a 6-vector of canonical coordinates and momenta. When using arc length as the independent variable, in Cartesian coordinates, $\zeta=(x,p_x,y,p_y,t,p_t)$. Here, $t$ denotes deviation in arrival time from a reference particle, and $p_t$ denotes the negative of the total energy deviation. (If time is used as the independent variable, then $t$ and $p_t$ are replaced by the usual canonical longitudinal variables are $z$ and $p_z$.) The particle evolution is governed by a symplectic map M that transforms $\zeta$ from some initial location in the beam line to a final location, $\zeta^f = M\zeta^i$. Since M is symplectic, the Jacobi matrix, $M(\zeta^i) = \partial \zeta^f / \partial \zeta^i$ is symplectic and satisfies $M^T J M = J$, where $J$ is the fundamental symplectic 2-form. Next we define the beam second moment matrix, $\Sigma$, with elements $\sigma_{ij} = <\zeta_i\ \zeta_j>$, where $<.>$ denotes ensemble averaging. In the linear approximation, the evolution of $\Sigma$ is given by $\Sigma^f = M\Sigma^i M^T$. In systems for which $\Sigma$ is 2x2 block diagonal, the determinant of each 2x2 block is equal to the 1D rms emittance in that phase plane. (Whether this emittance is normalized or unnormalized depends on the definition of $\zeta$, *i.e.*, if the canonical momenta are scaled by a constant design momentum, the emittance is unnormalized; if they are scaled by $mc$, the emittance is normalized.) It is well known that, in the uncoupled and uncorrelated case, the 1D canonical rms emittances are invariant under linear symplectic transformations. The generalization to the fully coupled and correlated case is as follows [12]:

Since $\Sigma$ is a real symmetric positive definite matrix, it follows from Williamson's Theorem that there exists a symplectic matrix A such that [13],

$$A\Sigma A^T = \Lambda = \mathrm{diag}(\lambda_1,\lambda_1,\lambda_2,\lambda_2,\lambda_3,\lambda_3), \quad (2)$$

with all the $\lambda_j > 0$. The quantities $\lambda_j^2$ are mean-squared *eigen-emittances* that generalize the 1D mean-squared emittances, $\varepsilon_j = <q_j^2><p_j^2> - <q_jp_j>^2$ to the fully coupled case. Though the matrix A is not unique, it is the case that, for any matrix A that transforms $\Sigma$ to Williamson normal form, the eigen-emittances are independent of A, up to a reordering. When the beam matrix has no correlations between the phase planes, the eigen-emittances correspond to the 1D rms emittances.

There are symplectic matrix routines for finding A and $\lambda_j$ [12]. Also, the eigenvalues of $J\Sigma$ are equal to $\pm i\lambda_j$. Therefore if only $\lambda_j$ are required, they can be computed using standard eigenvalue routines for real matrices. However, the matrix A is very useful, *e.g.*, for design purposes. Let $\Sigma$ denote a beam matrix at some intermediate location $s$ in a beamline, let $\Sigma^f$ denote a *desired* beam matrix at a final location $s^f$, and let M denote the linear transfer matrix between $s$ and $s^f$. Then,

$$\Sigma^f = (MA^{-1})\Lambda(MA^{-1})^T. \quad (3)$$

The above equation, which involves the known eigen-emittances $\lambda_j$ and transforming matrix A, and the known (desired) final $\Sigma$ matrix, mathematically determines what the transfer matrix M must be to achieve the design goals.

Along with the invariance of the eigen-emittances, there are two important inequalities: First, there is a *classical uncertainty principle* which states that, under linear symplectic dynamics, the product of the final mean-squared deviations in $q_j$ and $p_j$ for any plane must exceed, or at best equal, the smallest squared eigen-emittance, $\lambda^2_{min}$,

$$\langle q_j^2 \rangle \langle p_j^2 \rangle \geq \lambda^2_{\min} \qquad j=1,2,3. \quad (4a)$$

Second, there is a *minimum emittance theorem* which states that, under linear symplectic dynamics, the final mean squared emittance in any plane, must exceed or, at best, equal $\lambda^2_{min}$:

$$\varepsilon_j^2 \geq \lambda^2_{\min} \qquad j=1,2,3. \quad (4b)$$

The information provided by the classical uncertainty principle and the minimum emittance theorem is useful when designing a beam line to perform emittance manipulations because it sets lower limits on what one can hope to achieve.

The power of the eigen-emittance concept is that, if the correct eigen-emittances can be formed where the beam is generated at the cathode (by tailoring specific beam correlations), they are the only beam emittances that the beam can have if all correlations between the dimensions are removed, independent of the transport between the beam cathode and where these correlations are removed. This eliminates any concern about what the transport in between actually is, as long as higher order correlations do not grow sufficiently large to prevent the unwinding of the linear correlations. The demonstration of FBT transform in an RF photoinjector at Fermilab [14] has shown, at least in that case, that higher order correlations can be made sufficiently small to recover the eigen-emittances. In the case of an FBT, the magnetic field on the cathode changes the canonical angular momentum of the emitted particles, and hence changes the *transverse* eigen-emittances from their values without magnetic field. The magnetic field also introduces $x$-$p_y$ and $y$-$p_x$ correlations in the initial beam matrix. These correlations can be removed at the end of the FBT by appropriate beam optics resulting in a flat beam having asymmetric rms emittances [8]. The conservation of canonical angular momentum in the axisymmetric FBT beam line ensures that nonlinearities in the photoinjector do not spoil our ability to recover the eigen-emittances.

The same idea can be applied to systems that involve transverse-longitudinal eigen-emittance manipulations, by imposing longitudinal-transverse correlations in $\Sigma$ at the cathode surface and removing them downstream. There are several ways to introduce such correlations. Let $x$ denote a transverse coordinate on the photocathode surface. An $x$-$p_t$ correlation can be produced through spatial variation of the electron work function at the photocathode surface. An $x$-$p_t$ correlation can also be produced by spatial dependence of the photoinjector drive laser frequency, which can be achieved by reflecting a broadband laser pulse from a grating. An x-t correlation can be achieved by directing a short laser pulse at an angle (not normal) to the cathode surface. Alternatively, an $x$-$t$ correlation can be achieved in a beam whose propagation is normal to the cathode surface by using a number of schemes that produce pulse-front tilt in the laser beam (*i.e.* the laser pulse has the time delay depending on the transverse coordinate). In these last two cases, different areas of the cathode surface emit electrons at different times. All these schemes result in changing the initial eigen-emittances associated with $x$ and $t$ (leaving the $y$ eigen-emittance intact), with accompanying correlations introduced in the $x$ and $t$ planes of the beam matrix $\Sigma$. In this Letter we do not pursue the problem of finding the transfer matrix which diagonalizes $\Sigma$, which is shown elsewhere [15]. It is important to note, though, that the drifts and the skew quadrupoles in the conventional FBT optics are equivalent to chicanes and transversely deflecting RF cavities in a longitudinal-transverse eigen-emittance manipulation scheme, which makes the problem of designing the explicit beam line similar to the case of the conventional FBT.

Next, we study the general properties of schemes to introduce longitudinal-transverse correlations in the initial beam matrix. For this analysis we introduce correlations among two phase planes that we will denote $x$ and $t$. Since the third dimension remains uncorrelated with the other two, our analysis involves a 4x4 beam matrix. As noted above, invariance of the eigen-emittances is valid only for linear beam dynamics and does not account for certain important effects in the injector such as nonlinear space-charge forces or RF field curvature. These effects certainly affect performance of eigen-emittance manipulation systems but are not addressed here. Here we assume that the nonlinear forces are negligible in the photoinjector and consider the eigen-emittances downstream to be unchanged, as would be the case for low-charge beams [16].

In what follows we will use the word *intrinsic* to denote the electron bunch quantities in the absence of any externally imposed correlations (*i.e.* the intrinsic bunch is produced by a laser pulse that propagates normal to the cathode, the laser has no pulse-front tilt, the cathode has no position-dependent work function, *etc.*)

As a specific example, consider a bunch produced by a normally propagating laser and consider two cases, one without pulse-front tilt and one with pulse-front tilt. Suppose the transverse sizes of the two pulses are the same. Let $(x_0,t_0)$ and $(x,t)$ denote coordinates (transverse position on the cathode, arrival time of a photon at that position) in the non-front-tilted case and front-tilted case, respectively. These are related by,

$$x = x_0, \quad (5)$$
$$ct = ct_0 + x_0 \tan\theta, \quad (6)$$

where $\theta$ is the angle of the pulse-front tilt. The photocathode emission is proportional to the laser intensity at the photocathode surface. As a result, the coordinates of the emitted electrons in the intrinsic bunch and in the bunch created by the front-tilted laser are also related by (5) and (6), where $(x_0,t_0)$ correspond to the intrinsic bunch, and $(x,t)$ correspond to the bunch produced by the front-tilted laser.

Equations (5) and (6) are a special case of a more general transformation that introduces correlations in the bunch phase space at the cathode surface,

$$\varsigma = (I + \alpha L)\varsigma_0, \qquad L = \begin{bmatrix} 0 & B \\ A & 0 \end{bmatrix}, \quad (7)$$

where $\alpha$ is an arbitrary scalar, $A$ and $B$ are 2x2 matrices, and $L$ is a 4x4 cross-correlations matrix describing reshaping of the bunch phase space. Transform (7) results in the following beam matrix $\Sigma$ related to the intrinsic beam matrix $\Sigma_0$

$$\Sigma = (I + \alpha L)\Sigma_0 (I + \alpha L^T) \quad (8)$$

Typically the intrinsic beam matrix $\Sigma_0$ is diagonal. If that condition is not satisfied, one can always choose new coordinates in which the intrinsic beam matrix is diagonal. These new coordinates can be determined using the same eigen-emittance formalism described above. The introduced cross-correlations matrix $L$ in new coordinates would have the same form as in old coordinates since the intrinsic beam matrix $\Sigma_0$ does not have any transverse-longitudinal correlations. Therefore, the intrinsic beam matrix $\Sigma_0$ can be considered diagonal without loss of generality. For all the methods of introducing correlations studied so far, we have found that the determinant of the transformation $\varsigma_0 \to \varsigma$ is unity, i.e. $\det(I + \alpha L) = 1$. It can be shown that this constraint implies, in turn, the conditions $\det(AB)=\text{Tr}(AB)=0$.

The eigen-emittances of the modified beam matrix $\Sigma$ are the positive solutions of the characteristic equation $\det(J\Sigma - i\lambda I) = 0$. In what follows we will define new variables $(x,z)$ instead of $(x,t)$, where $z=ct$. Taking into account the form of the cross-correlation matrix $L$ described by Eq. (7), one finds the characteristic equation after some straightforward algebra,

$$\lambda^4 - \lambda^2\left(\varepsilon_{x0}^2 + \varepsilon_{z0}^2 + \alpha^2 Q + \alpha^4\left(\varepsilon_{x0}^2 \det B^2 + \varepsilon_{z0}^2 \det A^2\right)\right) = -\varepsilon_{x0}^2 \varepsilon_{z0}^2, \quad (9a)$$

$$Q = \sigma_{x0}^2 \sigma_{z0}^2 \left(a_{21}^2 + b_{21}^2\right) + \sigma_{x0}^2 \sigma_{pz0}^2 \left(a_{11}^2 + b_{22}^2\right) + \sigma_{px0}^2 \sigma_{z0}^2 \left(a_{22}^2 + b_{11}^2\right) + \sigma_{px0}^2 \sigma_{pz0}^2 \left(a_{12}^2 + b_{12}^2\right) \quad (9b)$$

Here $a_{ij}$ and $b_{ij}$ are the elements of the matrices $A$ and $B$, respectively; $\varepsilon_{x0}$ and $\varepsilon_{z0}$ are intrinsic emittances (the same as intrinsic eigen-emittances, since $\Sigma_0$ has no correlations); and $\sigma_{x0}^2$, $\sigma_{px0}^2$, $\sigma_{z0}^2$, $\sigma_{pz0}^2$ are the diagonal elements of $\Sigma_0$.

The biquadratic equation (9) defines two eigen-emittances which differ from the intrinsic emittances $\varepsilon_{x0}^2 = \sigma_{x0}^2 \sigma_{px0}^2$ and $\varepsilon_{z0}^2 = \sigma_{z0}^2 \sigma_{pz0}^2$ of the intrinsic beam matrix $\Sigma_0$. There are several important properties caused by the cross-correlation matrix $L$. First, the product of two eigen-emittances remains the same, which directly follows from the preservation of the phase space density, $\det(I + \alpha L) = 1$. Second, the eigen-emittance partitioning does not depend on the sign of the parameter $\alpha$ which reflects the symmetry of the intrinsic matrix $\Sigma_0$ which was assumed to be diagonal. Finally, the sum of two eigen-emittances grows with $\alpha$ since $Q>0$ and $\det A^2, \det B^2 \geq 0$. This property states that the maximum-to-minimum eigen-emittance ratio can *only increase* when cross-correlations in the beam matrix are introduced. Hence, these schemes cannot be used to create a bunch with equal emittances. At the same time, they can be used for reducing the smallest eigen-emittance which can be beneficial for producing electron beams with ultra high transverse brightness.

We now return to the specific case of a normally propagating, pulse-front-tilted laser beam. The quantities in Eq. 7 are,

$$A = \begin{bmatrix} 1 & 0 \\ 0 & 0 \end{bmatrix} \quad B = \begin{bmatrix} 0 & 0 \\ 0 & 0 \end{bmatrix} \quad \alpha = \tan\theta, \quad (10)$$

resulting in the following characteristic equation,

$$\lambda^4 - \lambda^2\left(\varepsilon_{x0}^2 + \varepsilon_{z0}^2\left(1 + \tan^2\theta \frac{\sigma_{x0}^2}{\sigma_{z0}^2}\right)\right) + \varepsilon_{x0}^2\varepsilon_{z0}^2 = 0. \quad (11)$$

When $\theta=0$, the solutions of Eq. (11) are the intrinsic emittances. Examination of Eq. (11) shows that the eigen-emittances can be changed significantly from the intrinsic values in the regime where $\tan^2\theta\, \sigma_{x0}^2/\sigma_{z0}^2 \gg 1 + \varepsilon_{x0}^2/\varepsilon_{z0}^2$. Stated differently, $\tan^2\theta \gg \sigma_{z0}^2/\sigma_{x0}^2 + \sigma_{px0}^2/\sigma_{pz0}^2$. For reasonable values of $\theta$, this condition is satisfied for a pancake-shaped laser pulse ($\sigma_{x0}^2 \gg \sigma_{z0}^2$). It is also satisfied for a bunch whose intrinsic energy spread is much greater than its transverse canonical momentum spread ($\sigma_{pz0}^2 \gg \sigma_{px0}^2$). In this regime the eigenemittances and their ratio can be approximated by,

$$\lambda_1 \sim \sigma_{x0}\sigma_{z'0}\tan\theta, \qquad \lambda_2 \sim \sigma_{z0}\sigma_{x'0}/\tan\theta. \quad (12)$$

$$\frac{\lambda_{\max}}{\lambda_{\min}}\frac{\varepsilon_{\min}^0}{\varepsilon_{\max}^0} \approx \tan^2\theta \frac{\sigma_{x0}^2}{\sigma_{z0}^2} \gg 1, \quad \varepsilon_{z0} \geq \varepsilon_{x0}. \quad (13)$$

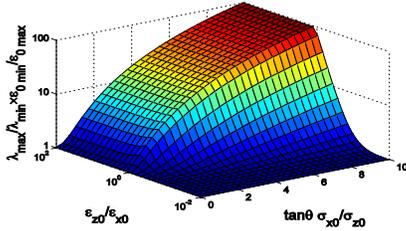

Fig. 1 (Color online) The change in the maximum-to-minimum eigen-emittance ratio by introducing pulse-front tilt in the laser pulse in the photoinjector.

In this regime it is therefore possible to produce large eigen-emittance asymmetry depending on the pulse-front tilt angle and the intrinsic rms quantities. This is confirmed in Fig. 1, which shows a plot of the maximum-to-minimum eigen-emittance ratio (normalized by the intrinsic ratio). On the other hand, in the opposite regime (a pencil-shaped laser pulse, or an emitted beam with $\sigma_{pz0}^2 \ll \sigma_{px0}^2$), the eigen-emittances do not significantly change for moderate values of the tilt angle $\theta$. At the same time, the electron bunch is reshaped in $x$-$z$ plane, which can result in smaller emittance growth due to nonlinear forces.

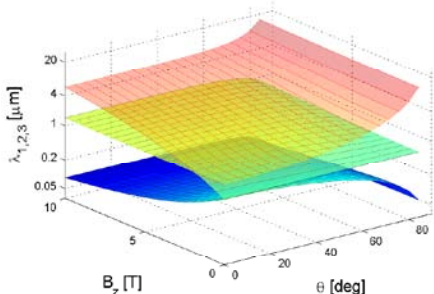

Fig. 2 (Color online) Partitioning of the eigen-emittances as both the axial magnetic field and laser pulse-front tilt angle are varied.

Next consider the longitudinal-transverse correlations that can be introduced with simultaneous application of an axial magnetic field at the cathode surface and a pulse-front tilted laser pulse. The logarithms of the eigen-emittances corresponding to such a system, for a beam having intrinsic emittances of 0.7/0.7/1.4 µm, are plotted in Fig. 2. Note that each of the eigen-emittances form clear surfaces that do not cross other than just in one point. As is evident from Fig. 2, for the intrinsic emittances mentioned, varying the magnetic field and pulse-front tilt do not produce the eigen-emittances of 0.14, 0.14, and 180 µm, that are required for this type of X-ray FEL. Other approaches are being analyzed with the eigen-emittance technique, such as use of an elliptical laser spot, to meet these eigen-emittance requirements.

The power of using eigen-emittances is that once generated, they are the only beam rms emittances that can occur, once correlations are removed, in a linear sense. The main issue to practical implementation of this technique is that this is a linear concept. It is critical that nonlinear correlations do not mask the linear correlations preventing them from being unwound. The Fermilab demonstration of the FBT [14] is a proof that, at least in that case, the nonlinear effects are small enough to be able to recover the eigen-emittances, with a significant emittance partitioning, even in a photoinjector (which, depending on the design, may involve significant nonlinear space-charge and RF effects). It is also worth noting that emittance compensation [17], by now a mature design technique, is based on the ability to mitigate and remove nonlinear correlations in photoinjectors. In addition, recent work has numerically demonstrated emittance compensation for an elliptical beam [18].

**References**
[1] P. Emma, *et al.*, *Nature Photonics* **4**, 6417 (2010).
[2] J. Rossbach, E. L. Saldin, E. A. Schneidmiller, and M. V. Yurkov, *Nucl. Instr. and Meth. A,* **374**, 401 (1996).
[3] http://marie.lanl.gov/ .
[4] H.-H. Braun, 48th ICFA Advanced Beam Dynamics Workshop on Future Light Sources, Menlo Park, 2010.
[5] A. Burov, S. Nagaitsev, and Y. Derbenev, *Phys. Rev. E* **66**, 016503 (2002).
[6] R. Brinkmann, Ya. Derbenev, and K. Floettmann, *Phys. Rev. ST Accel. Beams* **4**, 053501 (2001).
[7] K.-J. Kim, *Phys. Rev. ST Accel. Beams* **6**, 104002 (2003).
[8] B. E. Carlsten and K. A. Bishofberger, *New Journal of Physics* **8**, 286 (2006).
[9] M. Cornacchia and P. Emma, *Phys. Rev. ST Accel. Beams* **5**, 084001 (2002).
[10] P. Emma, Z. Huang, and K.-J. Kim, and Ph. Piot, *Phys. Rev. ST Accel. Beams* **9,** 100702 (2006).
[11] K.-J. Kim, presented at 13th Advanced Accelerator Concepts Workshop, Santa Cruz, 2008.
[12] A. Dragt *et al.*, *MARYLIE 3.0 Users' Manual*; A. Dragt, *Lie Methods for Nonlinear Dynamics with Applications to Accelerator Physics*, http://www.physics.umd.edu/dsat/ ; A. Dragt *et al.*, *Phys. Rev. A* **45**, 2572 (1992).
[13] J. Williamson, Amer. J. Math. **58**, 141 (1936).
[14] D. Edwards *et al.*, in *Proceedings of 2001 Particle Accelerator Conference, Chicago*.
[15] B. E. Carlsten, *et al.*, *to be submitted to Phys. Rev. ST Accel. Beams*.
[16] C. P. Hauri *et al.*, *Phys. Rev. Lett.* **104**, 234802 (2010).
[17] B. E. Carlsten, *Nucl. Instr. and Meth. A* **285,** 313 (1989).
[18] S.V. Miginsky**,** *Nucl. Instr. and Meth. A* **603,** 32 (2009).